\documentclass[conference,compsoc]{IEEEtran}

\usepackage[nocompress]{cite}  
\usepackage{amsmath,amssymb,amsfonts}
\usepackage{graphicx}
\usepackage{textcomp}
\usepackage{xcolor}
\usepackage{booktabs}
\usepackage{tabularx}
\usepackage{wasysym}        
\usepackage{url}            
\usepackage{hyperref}       
\usepackage{enumitem}       
\usepackage{dblfloatfix}    

\hypersetup{
  colorlinks=true,
  citecolor=blue,
  linkcolor=blue,
  urlcolor=blue,
}

\begin{document}

\title{ZERO-APT: A Closed-Loop Adversarial Framework for LLM-Driven Automated Penetration Testing under Intelligent Defense}

\author{
    \IEEEauthorblockN{Anlan Zheng}
    \IEEEauthorblockA{\textit{Zhejiang University of Technology}\\
    Hangzhou, China\\
    302023660008@zjut.edu.cn}
    \and
    \IEEEauthorblockN{Tiantian Zhu\textsuperscript{*}}
    \IEEEauthorblockA{\textit{Zhejiang University of Technology}\\
    Hangzhou, China\\
    \textsuperscript{*}ttzhu@zjut.edu.cn}
}
\maketitle

\begingroup
\renewcommand{\thefootnote}{\fnsymbol{footnote}}
\footnotetext[1]{Corresponding author.}
\endgroup

\begin{abstract}
LLM-driven automated penetration testing agents are typically evaluated against static targets that neither detect nor respond to attacks, so their behavior under intelligent defense remains untested. The causal consistency of multi-step attack chains likewise hinges on unstable LLM reasoning, and agent decisions remain opaque to human analysts. These three shortcomings, in realism, consistency, and auditability, are usually patched in isolation. We present ZERO-APT, a turn-based attacker-defender-judge framework that addresses them within a single architecture. For realism, ZERO-APT embeds a configurable LLM Defender that consumes Sysmon telemetry and detects attacks in real time, exposing the attacker to a live opponent rather than a passive target. For consistency, three architectural mechanisms move causal consistency from unstable LLM reasoning into enforced system architecture: separation of planning from execution, multi-dimensional ReAct feedback, and a hard-constraint-filtered action library. For auditability, a dedicated Judge agent adjudicates each round, maintains global state, and emits structured post-hoc CTI reports that make every decision traceable. We evaluate a Windows Server 2022 post-exploitation prototype across five scenarios with three Defender configurations. ZERO-APT reaches 79\% attack success rate (Aurora 22\%, PentestGPT 39\%), a Causal Consistency Score of 0.860 (Aurora 0.930, Claude Code 0.520), and end-to-end decision auditability through structured CTI reports. We release the benchmark to support evaluation of penetration agents under intelligent defense.
\end{abstract}


\section{Introduction}
\label{sec:introduction}

Penetration testing, a practice that simulates attacker behaviors through controlled intrusion to expose vulnerabilities, has long rested on manual expertise and remained labor-intensive and hard to scale~\cite{skandylas2025automated}. Large Language Models (LLMs) are now reshaping this practice at speed: more than 185 papers on AI-assisted security have appeared in top-tier venues through early 2025, with the curve steepening from 2023 onwards~\cite{xu2024llm}. Automated penetration testing is among the fastest-moving sub-areas, advancing from the first LLM-based privilege escalation demonstration~\cite{happe2023getting} to fully autonomous agent systems in 2024--2025~\cite{schroer2025sok}.

Three lines of work have organized this space. The first treats LLMs as auxiliary components and leans on PDDL or MCTS to enforce formal feasibility through hard preconditions~\cite{depasquale2024chainreactor,wang2026aurora,tung2026aegis}. The second formulates penetration as a Markov decision process and trains attack policies through massive simulated interaction~\cite{hu2020automated,li2023hierarchical}. The third places the LLM at the center of decision-making, as in PentestGPT's Reasoning-Generation-Parsing design~\cite{deng2024pentestgpt}, later augmented with retrieval and autonomous execution~\cite{shen2025pentestagent,xu2024autoattacker}. We pursue the third line: semantic understanding, contextual reasoning, and strategy generation are precisely the qualities a real adversarial environment demands of its planner~\cite{xu2024llm,schroer2025sok}. Three challenges, however, stand between this premise and a deployable system.

The first is \textbf{realism}. Existing attackers are almost always benchmarked against static targets that do not detect, respond, or counterattack~\cite{wang2026aurora,deng2024pentestgpt,shen2025pentestagent,xu2024autoattacker}, which marks a sharp departure from real adversarial settings, where every action leaves a detectable trace and the defender adjusts on the fly. To our knowledge, no published benchmark embeds a configurable LLM Defender as part of its evaluation infrastructure.

The second is \textbf{consistency}, specifically the causal consistency of an attack chain. Multi-step penetration requires each step's preconditions to be produced by earlier steps, yet LLMs are unstable over long horizons: hallucination, context forgetting, and logical drift repeatedly snap the chain~\cite{deng2024pentestgpt,evertz2026chasing}. Current remedies fall into two camps, with one confining the LLM to single-scenario exploitation~\cite{depasquale2024chainreactor,shen2025pentestagent,nakatani2025rapidpen} and the other replacing its decisions with classical planning or RL~\cite{wang2026aurora,li2025autonomous}. Neither generalizes to multi-step post-exploitation under adversarial pressure, and none confronts the issue at the architectural level.

The third is \textbf{auditability}. An LLM's internal chain of thought is opaque to outside analysts, and recent work shows these explanations are often unfaithful: models produce plausible-looking reasoning that hides the actual drivers of their decisions~\cite{turpin2023language}, and reasoning models in adversarial settings even conceal critical reasoning steps on purpose~\cite{anthropic2025reasoning}. Without auditable decisions, attack successes cannot be systematically understood, reproduced, or generalized to new environments.

ZERO-APT addresses these three concerns within a single attacker-defender-judge framework, with one architectural mechanism dedicated to each.

\textbf{Realism.} ZERO-APT introduces a configurable LLM Defender that detects attacks in real time from Sysmon telemetry through the ELK Stack~\cite{elastic2025elk}. The Defender spans three intensity tiers: L1 (weak perception, noisy logs), L2 (SOC-grade, Sigma-level signature rules combined with LLM reasoning), and L3 (known-adversary alerting enriched with TTP intelligence). Unlike static rule sets~\cite{le2021log,he2024llmelog}, the Defender acts as a tunable opponent, so an attacker is judged by what it does against an active adversary rather than against an inert target.

\textbf{Consistency.} Instead of asking the LLM to reason its way to a causally coherent attack chain, ZERO-APT lifts consistency out of the model and into the architecture, through three mechanisms detailed in \S\ref{sec:attacker}. (i) \textit{Planning-execution separation} splits high-level tactical decomposition from low-level micro-plan orchestration, blocking the cascade failures typical of monolithic ReAct loops~\cite{yao2023react,kung2023models}. (ii) \textit{Multi-dimensional ReAct feedback} extends the single intra-agent reasoning-action loop into two concurrent layers: one inter-agent (attacker-judge-attacker), one intra-agent (Executor self-debugging). (iii) \textit{Hard-constraint-filtered action library}: 775 post-exploitation actions, distilled from Atomic Red Team~\cite{redcanary2025atomic}, declare explicit preconditions that the system checks mechanically before execution, so feasibility is settled by the architecture rather than left to the LLM.

\textbf{Auditability.} A dedicated Judge agent adjudicates every round and maintains the global state, drives adaptive attack-path adjustment, and at session end emits a post-hoc CTI report aligned with STIX 2.0~\cite{oasis2021stix} and informed by prior work on attack-behavior alignment, threat-graph reconstruction, and APT representation~\cite{milajerdi2019poirot,cheng2025crucialg,aly2024megr}. What was once an opaque LLM decision process becomes a reviewable record that analysts can inspect, replay, and learn from.

We implement ZERO-APT for Windows Server 2022 post-exploitation and evaluate it on a custom benchmark of five scenarios, inspired by community ranges such as Metasploitable 3~\cite{rapid72016metasploitable} but designed to stress causal consistency, adaptive bypass, and backtrack recovery under three Defender tiers (L1--L3). Baselines are picked for progressive architectural comparison: Aurora~\cite{wang2026aurora} delivers consistency without realism, PentestGPT~\cite{deng2024pentestgpt} delivers preliminary interaction without hard constraints, and Claude Code~\cite{anthropic2025claudecode} delivers general-purpose reasoning without Judge closure or mechanical validation. ZERO-APT reaches 79\% end-to-end Attack Success Rate (ASR), ahead of Aurora at 22\% ($p<0.001$) and PentestGPT at 39\% ($p<0.01$), with a Causal Consistency Score (CCS) of 0.860 (Aurora 0.930, Claude Code 0.520). Aurora's high CCS paired with its lowest ASR shows that causal correctness alone does not survive an adaptive opponent. The ZERO-APT advantage in ASR also widens with Defender intensity, suggesting that its architectural mechanisms pay off most where defense bites hardest.

Our contributions are:
\begin{itemize}
\item We build the first attacker-defender-judge framework for adaptive attack chain generation, establishing a new evaluation paradigm in which an intelligent agent must defeat an active opponent rather than a static target.
\item We design three architectural mechanisms that move causal consistency from LLM reasoning into the system itself: planning-execution separation, multi-dimensional ReAct feedback, and a hard-constraint-filtered action library.
\item We release a three-tier configurable Defender that serves both as the core component of the first benchmark for penetration agents under intelligent defense and as a research platform for studying attacker-defender co-evolution.
\item We evaluate ZERO-APT across five scenarios on Windows Server 2022 with Sysmon telemetry, reporting gains over baselines in attack success rate, attack efficiency, and decision auditability.
\end{itemize}

\section{Related Work \& Motivation}
\label{sec:related}

\subsection{Related Work}

\textbf{Automated attack chain generation.} Three lines of research have shaped automated penetration testing~\cite{skandylas2025automated}, each making a distinct architectural bet on where intelligence should reside.

\textit{Classical planning with LLM assistance} places PDDL or MCTS at the reasoning core and reduces the LLM to a peripheral assistant~\cite{depasquale2024chainreactor,wang2026aurora,tung2026aegis}. Attack chains are precomputed against a known target description and dispatched in one shot, so a Defender alert mid-run cannot trigger replanning. The result is automation, not autonomy~\cite{mayoralvilches2025cybersecurity}.

\textit{Reinforcement learning (RL)} casts penetration as a Markov Decision Process (MDP) and trains policies through massive simulated interaction. Early DRL approaches~\cite{hu2020automated} have since been extended with hierarchical DRL~\cite{li2023hierarchical}, dynamic scenario adaptation~\cite{li2024dynpen}, knowledge-driven reward shaping~\cite{li2024knowledge}, and cross-topology generalization~\cite{terranova2024leveraging}. The recurring bottleneck is environmental overfitting~\cite{becker2024evaluation,terranova2024leveraging}: policies degrade against unfamiliar environments or intelligent defenders, because RL value functions do not grasp opponent intent in semantic terms.

\textit{LLM as planning core} has tracked LLM capability itself, evolving from code comprehension and tool use~\cite{happe2023getting,schick2023toolformer} and reasoning enhancement~\cite{wei2022chain} into systematic frameworks~\cite{deng2024pentestgpt}, autonomous execution~\cite{xu2024autoattacker,nakatani2025rapidpen,peng2025pwngpt,fang2024teams}, and multi-agent collaboration~\cite{shen2025pentestagent,kong2025vulnbot,muzsai2024hacksynth}. Stronger LLMs have produced stronger penetration agents~\cite{chan2025toward}, so the boundary of agent capability has so far been the boundary of model capability~\cite{tantakoun2025llm}. The system architecture, in turn, has been organized around squeezing more out of a single model. A different question rarely surfaces: in real adversarial settings with intelligent defense, can causal consistency and decision auditability be guaranteed by the architecture rather than left to LLM reasoning quality? ZERO-APT shifts the architectural focus from extending LLM capability to providing structural guarantees inside an attacker-defender loop.

\textbf{Automated defense and adversarial games.} On the attack side, all three lines treat defense as a static environmental feature~\cite{happe2023getting,depasquale2024chainreactor,wang2026aurora,deng2024pentestgpt,shen2025pentestagent,tung2026aegis,xu2024autoattacker,nakatani2025rapidpen}. On the defense side, surveys cover log-based endpoint detection~\cite{feng2025network,le2022log}, and recent work brings LLMs into detection pipelines: log anomaly detection augmented with threat intelligence~\cite{le2021log,he2024llmelog,guan2024logllm}, multi-host attack reconstruction from lightweight logs~\cite{liu2025musar}, multi-agent alert triage~\cite{wei2025cortex}, and provenance-based intrusion auditing~\cite{inam2023sok,yang2023prographer}. On the proactive side, hierarchical multi-agent RL supports collaborative defense decisions~\cite{singh2024hierarchical,tang2024method}, and LLM-MARL fusion sharpens defensive reasoning~\cite{xu2025l2m}. The two literatures, however, advance on separate tracks: attack research assumes no intelligent opponent, defense research targets known attack patterns~\cite{zhang2025when}, and we are not aware of any work that integrates both into a single closed-loop game. ZERO-APT embeds a configurable LLM Defender as a first-class system component, so the attacker is always evaluated against an opponent that both detects and attributes.

\subsection{Motivation}

The discussion above surfaces two structural shortcomings of existing work: each research line has its own bottleneck, and attack and defense research advance in isolation. Both observations remain coarse-grained, however, and do not, by themselves, determine ZERO-APT's architectural choices.

We build on the third line (LLM as planning core) because semantic understanding, contextual reasoning, and strategy generation are best suited to dynamic, open environments~\cite{xu2024llm,schroer2025sok}. LLM capability alone, however, has not been enough to deliver causal consistency~\cite{deng2024pentestgpt,evertz2026chasing}. Classical planning points to a remedy: Aurora~\cite{wang2026aurora} shows that architecture-level causal constraints can compensate for the LLM's reasoning gaps. RL contributes the loop structure: agents improve through try-feedback-adjust cycles~\cite{li2023hierarchical,li2024dynpen}, and their failure mode lies not in the loop itself but in RL's lack of semantic grasp of defender intent, which is precisely what LLMs do well.

The three observations point to a single architecture that fuses their strengths: LLMs as core decision-makers, architectural constraints to enforce causal consistency, and an internal adversarial feedback loop in place of RL's blind trial-and-error. To anchor this design, we compare PentestGPT~\cite{deng2024pentestgpt}, Aurora~\cite{wang2026aurora}, and Claude Code as three representative systems across realism, consistency, and auditability (Table~\ref{tab:comparison}).

\begin{table}[!t]
  \centering
  \caption{Architectural comparison of representative systems across three dimensions. \Circle=No, \CIRCLE=Yes, \LEFTcircle=Partial. Claude Code's three ratings are anchored in measured behavior, since it confronts no Defender, attains CCS=0.520 with Causal Break Failure rate (CBF\%)=27.4\% in our experiments, and emits unstructured execution traces rather than CTI reports.}
  \label{tab:comparison}
  \begin{tabular*}{\columnwidth}{@{\extracolsep{\fill}}lccc@{}}
    \toprule
    \textbf{System} & \textbf{Realism} & \textbf{Consistency} & \textbf{Auditability} \\
    \midrule
    PentestGPT~\cite{deng2024pentestgpt} & \Circle{} & \Circle{} & \Circle{} \\
    Aurora~\cite{wang2026aurora} & \Circle{} & \CIRCLE{} & \Circle{} \\
    Claude Code & \LEFTcircle{} & \Circle{} & \LEFTcircle{} \\
    \textbf{ZERO-APT} & \textbf{\CIRCLE{}} & \textbf{\CIRCLE{}} & \textbf{\CIRCLE{}} \\
    \bottomrule
  \end{tabular*}
\end{table}

The three gaps in Table~\ref{tab:comparison} are entangled, not independent. Without realism, evaluation exerts no adversarial pressure; without consistency, adding defense only widens causal breaks; without auditability, successful attacks cannot be understood or reproduced. Patching one gap in isolation does not work, because the missing dimensions undercut the patch. ZERO-APT therefore closes all three at once inside a single loop: the Defender supplies the realism pressure, which forces the Attacker to backtrack and so preserves consistency, while the Judge's global record delivers auditability. The three dimensions reinforce one another in a cycle in which the Attacker attempts an action, observes detection, backtracks, and retries.

\section{System Design}
\label{sec:design}

\subsection{Overview}

ZERO-APT runs the Attacker, Defender, and Judge as a turn-based loop. Each round proceeds Attacker $\rightarrow$ Defender $\rightarrow$ Judge: the Attacker performs an action, the Defender inspects system logs, and the Judge adjudicates the round and updates the global state, which then drives the next round's strategy. The loop continues until the mission goal is met or the per-scenario round limit (S1=8, S2=10, S3=12, S4=12, S5=9) is reached.

\begin{figure*}[t]
  \centering
  \includegraphics[width=\textwidth]{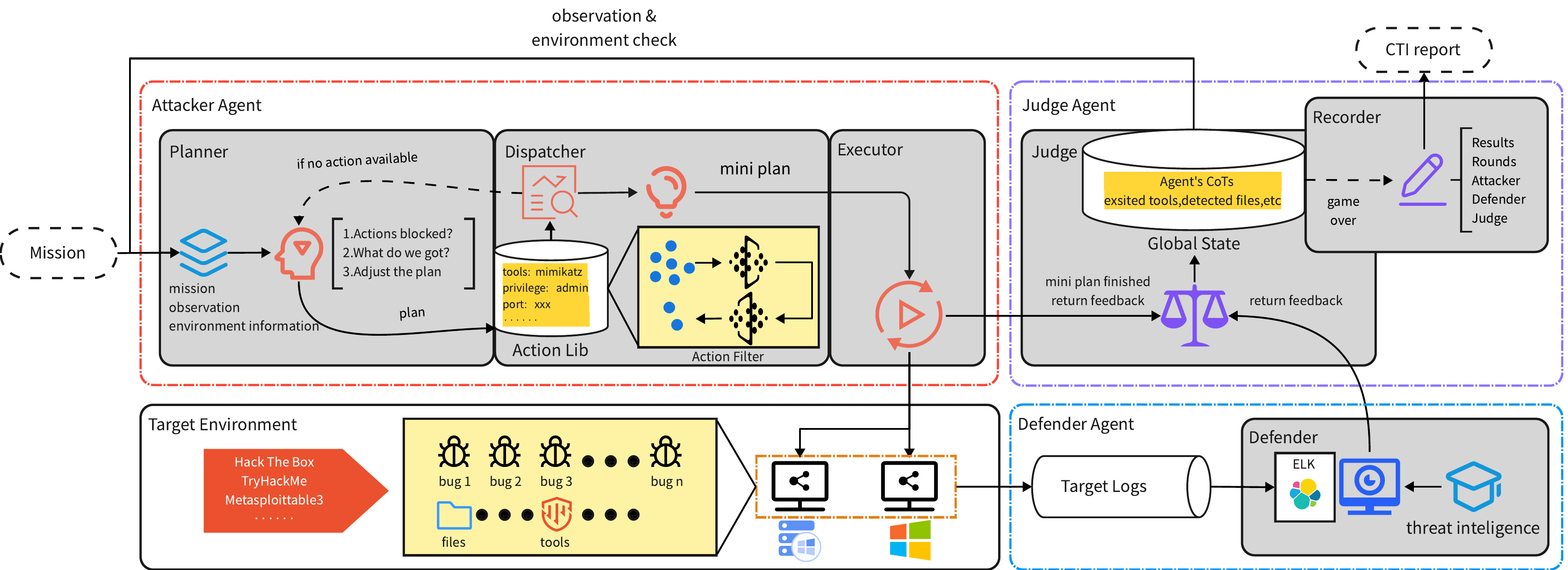}
  \caption{ZERO-APT overall architecture. Attacker executes actions and receives Judge feedback; Defender detects from Sysmon logs via ELK Stack; Judge adjudicates each round and drives the closed-loop game.}
  \label{fig:architecture}
\end{figure*}

The three agents map onto the three concerns introduced in Section~\ref{sec:introduction} (Figure~\ref{fig:architecture}). The Defender (Section~\ref{sec:defender}) runs an LLM-enhanced detection engine over Sysmon and the ELK Stack, isolated from the Attacker and reading only system logs. The Attacker (Section~\ref{sec:attacker}) enforces causal consistency through planning-execution separation, multi-dimensional ReAct feedback, and a hard-constraint-filtered action library. The Judge (Section~\ref{sec:judge}) decides each round's winner, updates the environment state, and produces a post-hoc CTI report. The three agents share a structured global state including all environmental information, but each agent reads only the slice it is authorized to access.

\subsection{Defender}
\label{sec:defender}

Existing attackers are benchmarked against static targets that perform no detection, no response, and no counteraction~\cite{wang2026aurora,deng2024pentestgpt,shen2025pentestagent,xu2024autoattacker}. ZERO-APT instead embeds an LLM Defender as a first-class system component, running a three-stage detection pipeline. The Monitor first pulls Sysmon~\cite{microsoft2025sysmon} events from the ELK Stack~\cite{elastic2025elk} within a 1-minute sliding window and formats them as structured text. Under L2 and L3, the events are then matched against four Sysmon signature rules over Event IDs 1/3/11/12--14 (Appendix~\ref{sec:defender_rules}); L1 skips this step. Finally, the matched events are annotated and fed to the Monitor LLM, which emits an intrusion judgment, a MITRE ATT\&CK~\cite{mitre2025attack} technique ID, a confidence score (0--100), an evidence summary, and an attack-intent rationale. The detection conclusion is written into global state for the Judge to adjudicate.

The Defender reads only system logs through the global state, and never the Attacker's internal plans or command sequences, mirroring real defense in which only externally observable traces are available.

Defender intensity is defined along two orthogonal dimensions: whether normal-activity noise is mixed into Sysmon logs, and whether attacker threat intelligence is injected into the Monitor LLM. The two dimensions cross to yield three progressive intensity tiers (Table~\ref{tab:defender_tiers}), with a shared baseline of four Sysmon signature rules underlying L2 and L3.

\begin{table}[!t]
  \centering
  \caption{Defender three-tier intensity design.}
  \label{tab:defender_tiers}
  \renewcommand{\arraystretch}{1.3}
  \begin{tabular*}{\columnwidth}{@{\extracolsep{\fill}}lcc@{}}
    \toprule
    & \textbf{No Threat Intel.} & \textbf{With Threat Intel.} \\
    \midrule
    \textbf{With noise} & L1 (Low) & --- \\
    \textbf{Clean logs} & L2 (Medium) & L3 (High) \\
    \bottomrule
  \end{tabular*}
\end{table}

L1 corresponds to a security-weak environment: no signature rules, noisy logs, and pure LLM reasoning. L2 reflects a routine enterprise SOC, where four signature rules supply detection signals while the Monitor LLM performs contextual reasoning and false-positive suppression over clean logs. L3 layers a known-adversary posture on top of L2, injecting attacker TTP intelligence and accumulated detection history. The configuration's effect surfaces as a monotonic ASR decline, L1 $>$ L2 $>$ L3 (Section~\ref{sec:realism}). The Defender is modular and can be swapped for a more sophisticated detector without touching the Attacker or the Judge. The pressure it exerts is what drives the architectural guarantees on the Attacker side: under real intelligent defense, a causally inconsistent attack chain quickly falls apart.

\subsection{Attacker}
\label{sec:attacker}

LLMs are unstable over long horizons: hallucination, context forgetting, and logical drift all snap causal chains in multi-step attacks~\cite{deng2024pentestgpt,evertz2026chasing}. Current remedies either confine the LLM to single-scenario exploitation~\cite{depasquale2024chainreactor,shen2025pentestagent,nakatani2025rapidpen} or replace its decisions with classical planning~\cite{wang2026aurora,li2025autonomous}; neither generalizes to multi-step post-exploitation under adversarial pressure. ZERO-APT instead lifts the burden of causal consistency off the LLM and onto the system itself, through three architectural mechanisms.

\subsubsection{Planning-Execution Separation}

The Attacker decouples tactical planning from action execution into two tiers, allocated to three LLM components with distinct responsibilities.

\textbf{Planner.} The Planner holds the attack's global view, taking mission objectives, current environment knowledge, Judge feedback, and Dispatcher backtrack requests as input, and emitting a high-level plan together with the current subgoal. It handles two backtrack types: \textit{passive} backtracking when Judge feedback shows the prior round was detected and blocked, and \textit{active} backtracking when the Dispatcher reports no feasible action for the current subgoal. Both types are detailed in Section~\ref{sec:attacker}'s backtracking subsection.

\textbf{Dispatcher.} The Dispatcher closes the gap between a high-level subgoal and its concrete execution, and operates in two phases. Phase~1 retrieves candidate actions from the action library, drops any actions on the Judge's forbidden list, and triggers a backtrack to the Planner if none remain. Phase~2 reads each surviving action's command template, parameter definitions, and pre/post-conditions, then sequences the actions against the current environment state, identifies parameters that still need runtime exploration, and emits the executable command sequence. The two-phase split keeps action selection (a discrete library lookup) cleanly separate from command synthesis (a parameter-binding step), so a faulty command template never contaminates the candidate set, and a missing parameter never blocks the entire chain.

\textbf{Executor.} The Executor receives the mini-plan and runs it step-by-step inside an internal ReAct loop, each step bounded by a retry cap. Unlike PentestGPT's monolithic ReAct, ZERO-APT's Executor has explicit step boundaries and fault tolerance: the LLM decides on its own whether to execute, mark a step complete, or skip after hitting the retry cap. Commands run remotely on the target Windows host via WinRM, with cross-host credential switching for lateral movement. Discovered file paths, harvested credentials, and confirmed environment characteristics flow back to the environment knowledge base in real time.

\subsubsection{Multi-dimensional ReAct Feedback}

ZERO-APT extends ReAct from a single intra-agent reasoning-action loop into two concurrent feedback layers, one inter-agent and one intra-agent.

\textbf{Inter-agent feedback: Attacker--Judge--Attacker closure.} Whenever the Defender detects an attack and the Judge rules for the Defender, the Planner receives the Judge's adjudication in the next round, carrying the Defender's detection basis and the ruling rationale, and revises the attack path accordingly. The Attacker no longer issues one-shot attempts; it adapts to a live opponent in real time.

\textbf{Intra-agent feedback: Executor self-debugging loop.} When a single command fails, the Executor inspects the error output, adjusts parameters, and retries; once the retry cap is hit, it skips the current step rather than blocking the run. The Executor advances on its own, so a transient error neither snaps the attack chain nor burns rounds on blind retries.

\subsubsection{Hard-Constraint-Filtered Action Library}

The third mechanism is mechanical validation at the system level: every action is checked against its declared conditions before being issued. The library holds 775 post-exploitation actions distilled from Atomic Red Team~\cite{redcanary2025atomic} and refined with LLM assistance, each carrying explicit pre- and postconditions (Table~\ref{tab:action_structure}).

\begin{table}[!t]
  \centering
  \caption{Attack action structure definition.}
  \label{tab:action_structure}
  \begin{tabular*}{\columnwidth}{@{\extracolsep{\fill}}llp{4.2cm}@{}}
    \toprule
    \textbf{Attribute} & \textbf{Type} & \textbf{Description} \\
    \midrule
    Action ID & string & Unique ID mapping to MITRE ATT\&CK technique (Txxxx) \\
    Platform & string & Target platform (Windows/Linux) \\
    Executor & string & Execution environment (PowerShell/CMD) \\
    Preconditions & object & Hard: min.\ privilege, tool deps, network ports \\
    Postconditions & object & Effects: created files, processes, credential types \\
    Command Template & string & Parameterized; parameters filled by Executor at runtime \\
    \bottomrule
  \end{tabular*}
\end{table}

\textbf{Hard--soft condition split and filtering.} Building on Aurora's Action-Aware Linking Model (AALM)~\cite{wang2026aurora}, ZERO-APT defines hard conditions as structured fields that the Dispatcher matches mechanically, without invoking a PDDL planner. PDDL encodes pre- and postconditions as logical predicates and runs a symbolic planner at runtime to search for a satisfying action sequence; ZERO-APT keeps Aurora's structured-condition declaration but matches conditions through Dispatcher table lookups, so no planner is needed. Hard conditions, including privilege levels, tool dependencies, and required ports, are matched against the environment state, while soft conditions such as file paths, account names, and configuration values are gathered by the Executor at runtime through exploration. Hard conditions default to \texttt{true}: the system optimistically assumes the target supports general capabilities. The Executor probes the actual state each round and flips any unmet condition to \texttt{false}; the next round's Dispatcher then excludes the actions that depend on it. The filter therefore tightens as exploration proceeds: every path is open at the start, exploration discovers the boundaries, and the action space converges to the feasible set. Because the six-tuple abstraction is platform-agnostic, extending the action library to Linux or container post-exploitation only requires new entries, since the Dispatcher's matching logic stays the same.

\subsubsection{Backtracking and Adaptive Exploration}

The Attacker's adaptive capability rests on the Planner's two backtrack paths. \textit{Passive backtracking} is cross-round and Judge-driven. When the Judge rules a round as failed due to environmental constraints or Defender detection, the Planner reads the Judge's observation in the next round and revises the attack path. \textit{Active backtracking} is same-round and Dispatcher-driven. When the Dispatcher finds no action satisfying its preconditions, it backtracks to the Planner, which redirects the subgoal, for instance from exploit-based escalation to tool-deployment escalation, subject to a per-round cap of three attempts.

\begin{figure*}[t]
  \centering
  \includegraphics[width=\textwidth]{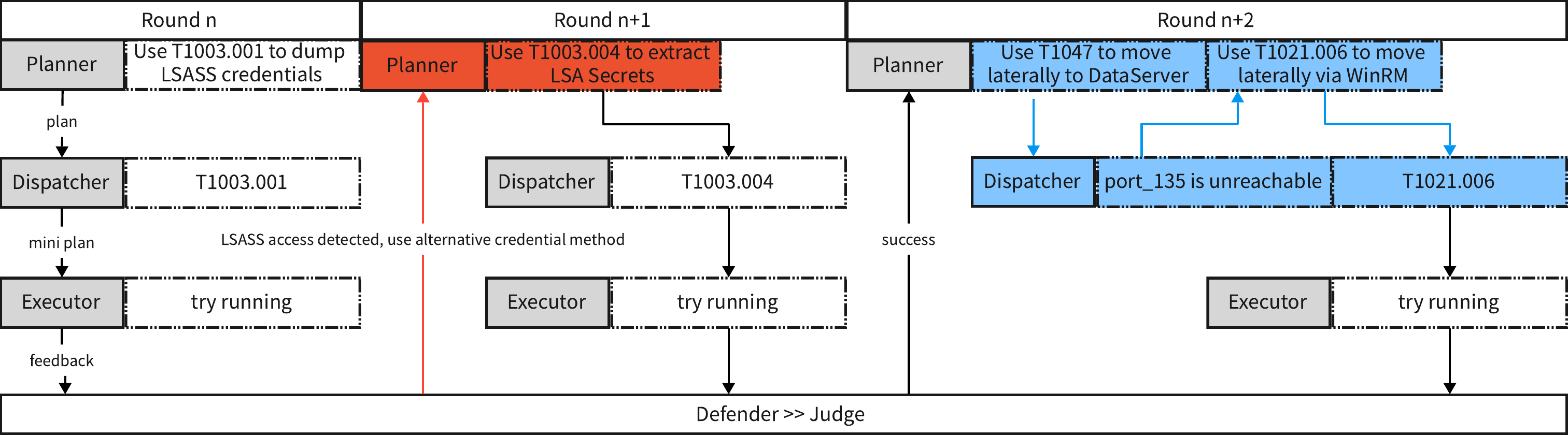}
  \caption{Two backtracking mechanisms. Red: passive backtracking, cross-round Judge-driven. Blue: active backtracking, same-round Dispatcher-driven.}
  \label{fig:backtracking}
\end{figure*}

Figure~\ref{fig:backtracking} illustrates both mechanisms in a credential extraction and lateral movement scenario. The red path traces a passive backtrack. SG1 (T1003.001) is detected in Round~$n$, the Judge rules Blue Win, and in Round~$n+1$ the Planner backtracks and replaces SG1 with T1003.004, which executes successfully. The blue path traces an active backtrack. The precondition check on SG2 (T1047) finds port 135 unreachable, so the Planner backtracks within the same round and switches to T1021.006 (WinRM), which executes successfully.

\subsection{Judge}
\label{sec:judge}

Existing automated penetration evaluations decide attack success post hoc, either through manual assessment or through environment-defined termination conditions~\cite{happe2025benchmarking,isozaki2025towards}. Neither approach suits a turn-based real-time game, in which an action may execute successfully yet still be exposed at detection. Online per-round adjudication is needed instead. Drawing on the Controller concept in multi-agent penetration~\cite{geng2025controller}, ZERO-APT promotes this role to a first-class Judge with independent adjudication authority and post-hoc audit capability. The Judge carries two duties, per-round winner adjudication and post-hoc CTI report generation (Summary).

\subsubsection{Per-Round Adjudication and State Management}

The Judge operates online after each Attacker--Defender exchange. The Judge LLM receives both the Attacker's complete round feedback (command sequence, exploration results, execution analysis) and the Defender's structured detection output, then scores the Defender's evidence quality on a 0--100 scale based on its correspondence to actual Attacker behavior and its logical sufficiency. Adjudication uses a threshold of 75 (Table~\ref{tab:judge_logic}).

\begin{table*}[t]
  \centering
  \caption{Judge adjudication logic.}
  \label{tab:judge_logic}
  \begin{tabular*}{\textwidth}{@{\extracolsep{\fill}}lcll@{}}
    \toprule
    \textbf{Condition} & \textbf{Verdict} & \textbf{Consequence for Attacker} & \textbf{State Update} \\
    \midrule
    evidence\_score $\geq$ 75 & Defender wins & Round's attack results not recognized & Blocked actions added to forbidden list \\
    evidence\_score $<$ 75 & Attacker wins & Attack results written to state & Discovered files, credentials, etc.\ enter knowledge base \\
    \bottomrule
  \end{tabular*}
\end{table*}

Regardless of outcome, the Judge outputs a natural-language ruling, and this ruling is the sole channel through which defense-side information reaches the Attacker. When the Attacker wins, the ruling points out the Defender's blind spots. When the Defender wins, it records the detection evidence and attribution, which the Planner reads in the next round to revise its strategy and evade similar detection.

The 75-point threshold is calibrated to the three Defender tiers. Under L1, with noisy logs and no signature rules, Defender confidence rarely reaches 75 and the Judge typically rules for the Attacker. Under L2, signature matches anchor the LLM's reasoning and lift confidence into a contested band around the threshold. Under L3, layered TTP intelligence pushes confidence higher still and crosses the 75-point line most often. The TP/FP data across tiers (Section~\ref{sec:realism}, Figure~\ref{fig:defender_perf}) backs this calibration, with stronger tiers triggering the adjudication threshold more frequently.

\subsubsection{Post-Hoc CTI Report Generation}

The Summary is the Judge's offline component, executing once after experiment termination. It takes the complete interactive history as input and emits a structured CTI report organized into two layers. The historical layer records the attack overview, the per-round timeline, and the evolution of the forbidden action list. The analytical layer assesses Attacker strategy under Defender pressure, locates the Defender's systematic blind spots, and grounds defensive recommendations in them. Attacker reasoning chains, Defender detection logic, Judge adjudication basis, and backtrack trigger reasons are all structurally recorded, giving analysts a fully traceable decision chain.

\section{Evaluation}
\label{sec:evaluation}

We conduct 1,400 experiments (5 scenarios $\times$ 4 schemes $\times$ 3 Defender intensities $\times$ 20 repetitions = 1,200, plus 5 scenarios $\times$ 2 ablation variants $\times$ 20 repetitions = 200) to test ZERO-APT's fulfillment of its three design goals: realism, consistency, and auditability.

\subsection{Experimental Setup}

\textbf{Why a custom benchmark?} Existing penetration benchmarks split into two camps, neither of which pits an attacker against intelligent defense. Static ranges such as HackTheBox~\cite{hackthebox2025}, VulnHub~\cite{vulnhub2025}, CTF competitions~\cite{shao2024nyu}, and recent LLM penetration frameworks~\cite{yang2025pentesteval,gioacchini2025autopenbench,zhu2025cve} expose realistic vulnerabilities but carry no LLM-based real-time detection or attribution~\cite{happe2025benchmarking}. Autonomous agents reach about 60\% on Easy targets, near zero on Hard~\cite{deng2024pentestgpt}, and 21--31\% on broader benchmarks~\cite{yang2025pentesteval,gioacchini2025autopenbench}, all measuring attack capability without detection pressure. Retrospective attack-defense datasets such as DARPA TC and OpTC~\cite{inam2023sok} record real attack logs but stay offline, so they cannot host the closed-loop game we need.

We therefore build a custom benchmark on top of community ranges such as Metasploitable 3~\cite{rapid72016metasploitable}, with the target updated to Windows Server 2022 to match the Atomic Red Team~\cite{redcanary2025atomic}-derived action library. Five scenarios (S1--S5), each parameterized by intelligent defense level (L1--L3), stress one architectural capability at a time. Aurora~\cite{wang2026aurora} and PentestGPT~\cite{deng2024pentestgpt} likewise evaluate on custom scenarios~\cite{happe2025benchmarking}.

\textbf{Test environment.} Two Windows Server 2022 VMs: primary target ZEROAPTtarget (192.168.32.139, domain controller) and lateral movement target ZEROAPTtarget-DataServer (192.168.1.102, separate subnet, S3). Both deploy ELK Stack for Sysmon event log collection. The Attacker executes commands via Windows Remote Management (WinRM); the Defender queries \texttt{winlogbeat-*} through Elasticsearch; the Judge independently runs adjudication and report generation. All LLM modules use Claude Opus 4.7~\cite{anthropic2025claude}.

\textbf{Five post-exploitation scenarios.} The five scenarios (Table~\ref{tab:scenarios}) each target a different facet of ZERO-APT. S1 (Credential Theft) runs a basic multi-step causal chain under standard detection. S2 (Persistence + Privilege Escalation) adds multi-branch path selection and stresses hard-constraint filtering against more complex causal dependencies. S3 (Adversarial Lateral Movement) lifts the Defender to L3 and forces cross-subnet adaptive bypass. S4 (Covert Data Exfiltration) stretches the chain to four stages, where a single detection compromises the mission. S5 (Path Blocking Adaptation) holds the Defender at L1 but blocks the primary path upfront, so success rides entirely on backtrack recovery rather than detection avoidance.

\begin{table}[!t]
  \centering
  \caption{Five adversarial post-exploitation scenarios.}
  \label{tab:scenarios}
  \renewcommand{\arraystretch}{1.35}
  \renewcommand{\tabcolsep}{3pt}
  \begin{tabular*}{\columnwidth}{@{\extracolsep{\fill}}c>{\raggedright\arraybackslash}p{1.7cm}c>{\raggedright\arraybackslash}p{1.9cm}>{\raggedright\arraybackslash}p{2.5cm}@{}}
    \toprule
    \textbf{Sc.} & \textbf{Name} & \textbf{Def.} & \textbf{Tested Capability} & \textbf{Key Adversarial Feature} \\
    \midrule
    S1 & Credential Theft & L2 & Causal chain execution & Multi-step causal chain under standard detection \\
    S2 & Persistence + Privesc & L2 & Multi-branch path selection & Hard-constraint filtering under complex dependencies \\
    S3 & Lateral Movement & L3 & Adaptive bypass & L3 full TTP detection; cross-subnet movement \\
    S4 & Covert Exfiltration & L2 & Stealth long-chain & Four-stage chain; detection = mission compromise \\
    S5 & Path Blocking & L1 & Backtrack recovery & Primary path pre-blocked; Attacker must reroute \\
    \bottomrule
  \end{tabular*}
\end{table}

\textbf{Comparison schemes.} The three baselines pick out three distinct architectural choices (Section~\ref{sec:related} Table~\ref{tab:comparison} gives the full comparison). Aurora~\cite{wang2026aurora} guarantees causal consistency through PDDL but generates attack scripts statically and does not respond to Defender feedback. PentestGPT~\cite{deng2024pentestgpt} in human-in-the-loop mode introduces preliminary closed-loop interaction through Pentesting Task Trees (PTTs), without hard-constraint filtering or structured backtracking. We exclude PentestGPT v1.0~\cite{greydgl2025pentestgpt} because it wraps an externally attached Claude Code, mixing its own contribution with that of the wrapped agent. Claude Code itself enters as a standalone baseline for strong general-purpose LLM reasoning without domain architecture, Judge closure, or mechanical precondition validation.

\begin{itemize}[leftmargin=*]
\item \textbf{ZERO-APT}: Complete system; Defender configured per-scenario as L1/L2/L3.
\item \textbf{Aurora~\cite{wang2026aurora}}: Action library preconditions/postconditions translated to PDDL; attack scripts generated once, no Defender feedback, no execution-time path adjustment.
\item \textbf{PentestGPT~\cite{deng2024pentestgpt}}: Human-in-the-loop mode. PentestGPT generates commands; the human executes via WinRM and feeds results back without making attack decisions. Receives Defender detection results as environment feedback; does not access ZERO-APT action library.
\item \textbf{Claude Code}: Provided command execution interface; autonomously writes and executes commands via WinRM. Can reference action library without hard-constraint filtering (preconditions self-assessed by LLM); Judge feedback contains only success/failure.
\end{itemize}

\textbf{Default scenario-Defender pairing.} Each scenario carries a default Defender level (S1/S2/S4=L2, S3=L3, S5=L1) used for the consistency, command-effectiveness, and ablation analyses in Section~\ref{sec:consistency}. The pairing tracks how enterprise Security Information and Event Management (SIEM) is actually deployed, where different network zones face very different detection intensities~\cite{uetz2023you,le2021log,he2024llmelog}. S3 at L3 captures a known-adversary high-alert posture, S5 at L1 isolates the environmental variable for backtrack recovery testing, and S1/S2/S4 at L2 mirror a routine enterprise SOC.

\textbf{Full-tier realism validation.} To check that the Defender imposes real measurable pressure, we run every scheme on every scenario at every Defender level. If escalation suppresses all attackers systematically, ASR should fall monotonically, and the most architecturally resilient scheme should decay slowest (Section~\ref{sec:realism}).

\textbf{Defender intensity construction.} L2 and L3 share four Sysmon signature rules as the detection baseline (Section~\ref{sec:defender}). Two orthogonal switches yield three progressive tiers (Section~\ref{sec:defender} Table~\ref{tab:defender_tiers}): whether Sysmon logs carry normal-activity noise, and whether the Monitor LLM receives attacker TTP descriptions. Prior work backs the effects of both noise and intelligence injection~\cite{le2021log,he2024llmelog,xu2025l2m}.

\textbf{Ablation variants.} Two ablation versions test independent contributions of architectural mechanisms.
\begin{itemize}[leftmargin=*]
\item \textbf{-ActionLib}: Remove action library; LLM generates commands from memory, retaining multi-layer planning + Judge closure. Tests whether LLM can maintain causal consistency without mechanical precondition validation.
\item \textbf{-MultiReAct}: Merge Planner + Dispatcher + Executor into single-layer ReAct, remove Judge closure, retain action library. Tests whether adaptive bypass capability vanishes when the system degrades to monolithic ReAct.
\end{itemize}

\textbf{Evaluation metrics.} Four core metrics anchor the analysis: three correspond to the three challenges and BSR isolates the architecture's adaptive recovery; auxiliary metrics are in the Appendix.
\begin{itemize}[leftmargin=*]
\item \textbf{ASR} (Attack Success Rate) = N\_success / N\_total, per-scenario and per-Defender-intensity.
\item \textbf{CCS} (Causal Consistency Score) = Two human security experts independently review each experiment's complete attack chain (command sequences, execution outputs, state transitions), starting from 10 points, deducting 1 per causal inconsistency, averaging and normalizing to $[0,1]$. Intraclass Correlation Coefficient (ICC)=0.84. Unlike mechanical precondition matching, expert scoring identifies implicit causal breaks. Collected on default scenarios + ablation experiments (600 total).
\item \textbf{CBF\%} (Causal Break Failure Rate) is the share of executed commands that fail because their causal preconditions are unmet, excluding commands that execute successfully but are caught by the Defender. Since every command is generated by Claude Opus 4.7 and contains virtually no syntax error, a non-zero exit code reflects an unmet upstream precondition rather than command malformation. CCS and CBF\% are complementary: CCS measures, from the positive side, whether causal conditions are met; CBF\% measures, from the negative side, whether causal breaks translate into execution failures. Collected on default scenarios plus ablations (600 in total).
\item \textbf{BSR} (Backtrack Success Rate) is the fraction of backtracks that recover into a successful subgoal, capturing the architecture's adaptive recovery capability under Defender pressure.
\end{itemize}

\begin{figure}[t]
  \centering
  \includegraphics[width=\columnwidth]{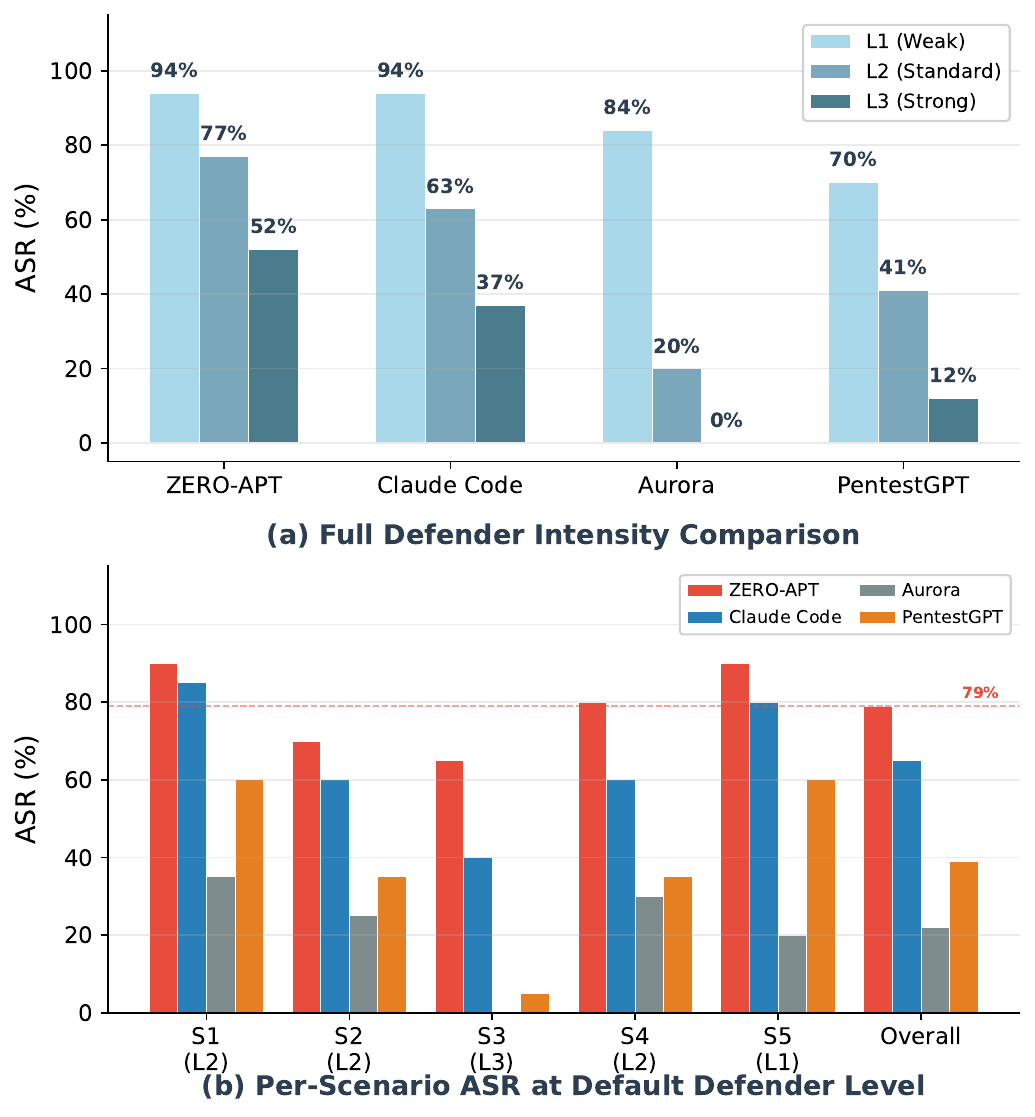}
  \caption{Attack success rate by Defender intensity. (a) ASR aggregated by scheme and Defender level (n=100 per bar). All schemes show monotonic L1$>$L2$>$L3 decline; ZERO-APT maintains highest ASR at every level; gap with baselines is largest at L3. (b) Per-scenario ASR under default Defender pairing (S1/S2/S4=L2, S3=L3, S5=L1; n=20 per bar).}
  \label{fig:asr_combined}
\end{figure}

\subsection{Realism Validation}
\label{sec:realism}

We test all four schemes on all five scenarios across three Defender intensity tiers. Figure~\ref{fig:asr_combined}(a) shows aggregated ASR (n=100 per bar); Table~\ref{tab:per_scenario_asr} gives the complete per-scenario per-intensity results (n=20 per cell).

\begin{table*}[t]
  \centering
  \caption{Per-scenario ASR by Defender level. Each (method, scenario, Defender intensity) cell runs 20 repetitions. Column headers show each scenario's default Defender pairing in parentheses. L1/L2/L3 defined in Section~\ref{sec:defender} Table~\ref{tab:defender_tiers}.}
  \label{tab:per_scenario_asr}
  \begin{tabular*}{\textwidth}{@{\extracolsep{\fill}}lccccc@{}}
    \toprule
    & \textbf{S1 (L2)} & \textbf{S2 (L2)} & \textbf{S3 (L3)} & \textbf{S4 (L2)} & \textbf{S5 (L1)} \\
    \midrule
    ZERO-APT L1 & 20/20 & 18/20 & 19/20 & 19/20 & 18/20 \\
    ZERO-APT L2 & 18/20 & 14/20 & 16/20 & 16/20 & 13/20 \\
    ZERO-APT L3 & 14/20 & 8/20 & 13/20 & 11/20 & 6/20 \\
    Claude Code L1 & 20/20 & 20/20 & 19/20 & 19/20 & 16/20 \\
    Claude Code L2 & 17/20 & 12/20 & 14/20 & 12/20 & 8/20 \\
    Claude Code L3 & 12/20 & 6/20 & 8/20 & 6/20 & 5/20 \\
    Aurora L1 & 20/20 & 20/20 & 20/20 & 20/20 & 4/20 \\
    Aurora L2 & 7/20 & 5/20 & 1/20 & 6/20 & 1/20 \\
    Aurora L3 & 0/20 & 0/20 & 0/20 & 0/20 & 0/20 \\
    PentestGPT L1 & 17/20 & 14/20 & 13/20 & 14/20 & 12/20 \\
    PentestGPT L2 & 12/20 & 7/20 & 8/20 & 7/20 & 7/20 \\
    PentestGPT L3 & 5/20 & 2/20 & 1/20 & 2/20 & 2/20 \\
    \bottomrule
  \end{tabular*}
\end{table*}

\textbf{Overall trends.} Figure~\ref{fig:asr_combined}(a) reads off a monotonic L1 $>$ L2 $>$ L3 decline across all four schemes (Friedman~\cite{friedman1937use} $\chi^2=8.40$, $p=0.015$), so escalating the Defender imposes systematic, measurable pressure on which the rest of the analysis rests. ZERO-APT decays most slowly, from 94\% at L1 to 77\% at L2 and 52\% at L3, a 42-point drop overall. Claude Code ties ZERO-APT at L1 (both 94\%) where general reasoning is enough against a near-blind Defender (TP=3.3\%), then falls behind by 14 points at L2 and 15 points at L3. The widening gap places ZERO-APT's architectural mechanisms exactly where defense bites hardest. L1 parity followed by L2/L3 divergence also rules out a benchmark-bias account, since all schemes share Claude Opus 4.7 and a biased benchmark would tilt every level alike rather than only those under detection pressure.

\textbf{Scenario-level analysis.} Table~\ref{tab:per_scenario_asr} reveals internal structure beneath the aggregate trend.

Aurora's failure mode is the most diagnostic. At L1 it goes near-perfect on S1--S4 (80/80) yet collapses on S5 (4/20), because the PDDL planner locks every path at planning time and a pre-blocked entry leaves no room for runtime rerouting. L2 admits only 20 successes out of 100, since signature rules quickly fingerprint the same static script. L3 with full TTP intelligence sends every scenario to zero. Causal correctness, baked in once, becomes a liability the moment the environment moves.

Claude Code turns sharper. L1 ties ZERO-APT at 94\% and even edges ahead on S2 (20/20 vs.\ 18/20), but L2 and L3 fall to 63\% and 37\% once self-correction has to substitute for structured backtracking and hard-constraint filtering. The same parity-then-divergence pattern reappears, so a stronger-base-model story does not fit ZERO-APT's gains either.

ZERO-APT under L3 still holds above 13/20 on two scenarios. S1 (14/20) leans on a short causal chain that the action library can re-route within a single category. S3 (13/20) turns heavier detection into more frequent Judge-driven backtracking, with each Defender block read off as a precise path exclusion. S4 falls to 11/20 once the four-stage chain stretches the exposure window. S2 (8/20) and S5 (6/20) sit lowest, where multi-branch dependencies and path pre-blocking compound with detection pressure.

PentestGPT reaches only 70\% even at L1, where the PTT tree carries no hard-constraint validation, and slips to 12/100 at L3. Its human-in-the-loop mode treats Defender feedback as environmental observation rather than a replanning signal, so under L3's joint detection-plus-intelligence pressure a single sub-node failure typically ends the whole chain instead of triggering a path switch.

All three ASR gaps widen as the Defender hardens. ZERO-APT pulls away from Aurora 10pp $\rightarrow$ 52pp (L3 zero), from PentestGPT 24pp $\rightarrow$ 40pp, and from Claude Code 0pp $\rightarrow$ 14pp at L2 and 15pp at L3.

\begin{figure}[t]
  \centering
  \includegraphics[width=\columnwidth]{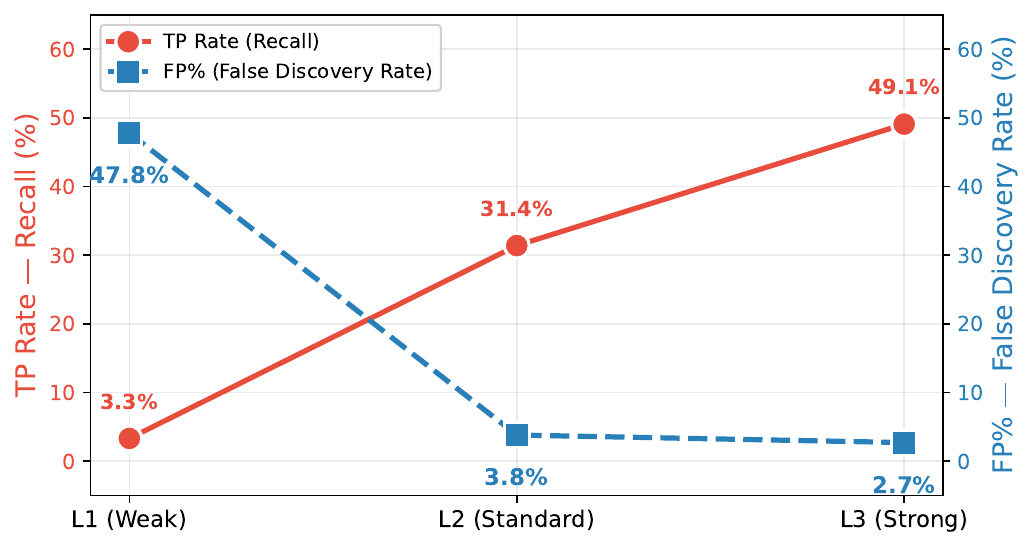}
  \caption{Defender detection performance across three intensities (ZERO-APT, n=100 per level). TP Rate monotonically rises from 3.3\% (L1) to 49.1\% (L3); FP\% drops from 47.8\% (L1) to 3.8\% (L2) and 2.7\% (L3), showing detection pressure derives from effective detection rather than blind alerting.}
  \label{fig:defender_perf}
\end{figure}

\textbf{Defender credibility.} One alternative reading is that the Defender depresses ASR through blind alerting rather than real detection. Figure~\ref{fig:defender_perf} reports TP Rate and FP\% across the three intensities to settle this. The Defender emits a binary intrusion call per round, and the Judge scores it whenever an intrusion is claimed; silent rounds drop out of both rates.

Figure~\ref{fig:defender_perf} rules out the blind-alert account. L2 and L3 FP\% land at 3.8\% and 2.7\%, while L1's 47.8\% traces to LLM misjudgments in a noisy environment. TP Rate climbs from 3.3\% at L1 to 31.4\% at L2 and 49.1\% at L3, mirroring the ASR decline. Attacker pressure thus comes from real detection, not random alerting. Almost all L1 true positives sit on S5, whose 4--6 rounds of backtracking widen the Defender's window; the shorter 2--3 round chains in S1--S4 yield essentially none.

L2 is not a pure Sigma baseline, because a Monitor LLM contextual reasoning layer sits on top of the four signature rules. Its 31.4\% TP and 3.8\% FP therefore reflect Sigma coverage and LLM reasoning together. A pure Sigma deployment in the same environment would match the TP, since community rule libraries cover broader behavior, but would post a higher FP because no LLM layer suppresses noise-driven false alerts. L1$\rightarrow$L2 jointly buys TP $+$28.1pp and FP $-$44.0pp from rule coverage plus LLM reasoning on cleaner logs; L2$\rightarrow$L3 isolates TTP intelligence at TP $+$17.7pp, no FP cost.

Across baselines, the TP Rate ranking inverts the ASR ranking exactly: Aurora $\sim$72\% / PentestGPT $\sim$40\% / Claude Code $\sim$32\% / ZERO-APT 31.4\%, against ASR ZERO-APT 77\% / Claude Code 63\% / PentestGPT 41\% / Aurora 20\%. The pattern matches ZERO-APT's minimum-exposure plus backtrack-recovery strategy: the Defender catches attackers that leak more surface area and flags architecturally more complete systems less often, one more sign that detection pressure is real rather than artificial.

\subsection{Consistency Validation}
\label{sec:consistency}

All analyses below use each scenario's default Defender configuration (S1/S2/S4=L2, S3=L3, S5=L1), so that Defender intensity matches the scenario's test goal.

\begin{figure}[t]
  \centering
  \includegraphics[width=\columnwidth]{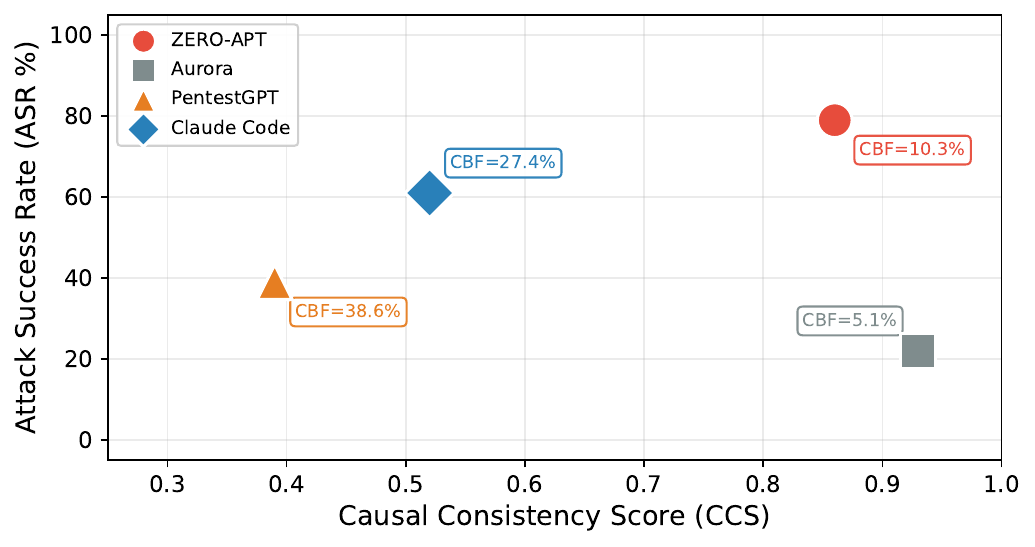}
  \caption{CCS-ASR scatter plot with CBF\% annotations. Distribution of four schemes in CCS-ASR two-dimensional space, each point annotated with CBF\%. Aurora has the highest CCS (0.930) but lowest ASR (22\%); ZERO-APT is the only system simultaneously achieving high CCS (0.860) and high ASR (79\%); PentestGPT occupies bottom-left (low CCS, low ASR); Claude Code in intermediate region.}
  \label{fig:ccs_scatter}
\end{figure}

Figure~\ref{fig:ccs_scatter} reads the four schemes across CCS, ASR, and CBF\%. Aurora's CCS (0.930) tops the chart, with PDDL formal constraints leaving expert reviewers almost no causal inconsistency to flag. ZERO-APT follows at 0.860 (vs.\ PentestGPT $p<0.001$, vs.\ Claude Code $p<0.001$), so the three architectural mechanisms lift an LLM-driven system close to formal-method-level consistency. Aurora's 0.930 paired with its 22\% ASR demonstrates that while causal consistency is necessary, it is not sufficient on its own. Indeed, chains stay causally correct yet die at the first adaptive demand. Only ZERO-APT lives in both the high-CCS and high-ASR region.

CBF\% tells the same story from the negative side. ZERO-APT records 10.3\%, well below Claude Code (27.4\%, $p<0.001$) and PentestGPT (38.6\%, $p<0.001$). Aurora's 5.1\% sits lowest because PDDL hard constraints rule out causally broken paths at planning time, an architectural guarantee outside the LLM-driven comparison. Among LLM-driven schemes, ZERO-APT alone pairs the highest CCS (0.860) and ASR (79\%) with a CBF\% within 5pp of the hard-constraint baseline.

Figure~\ref{fig:ccs_scatter} maps the four schemes onto the CCS-ASR plane. Aurora lands in the causally-correct-yet-inert quadrant; PentestGPT and Claude Code, lacking hard constraints and structured backtracking, scatter across the low-CCS or intermediate band; ZERO-APT, fusing hard-constraint filtering with adversarial feedback, alone occupies the high-CCS, high-ASR, low-CBF\% corner.

\textbf{Architecture ablation experiments.} Removing the action library (-ActionLib) and merging the multi-layer architecture (-MultiReAct) each caused systematic performance degradation (Table~\ref{tab:ablation}).

\begin{table}[!t]
  \centering
  \caption{Architecture ablation experiment results (n=100). -ActionLib removes the action library; LLM generates commands from memory. -MultiReAct merges three layers into single ReAct and removes Judge closure.}
  \label{tab:ablation}
  \begin{tabular*}{\columnwidth}{@{\extracolsep{\fill}}lccccc@{}}
    \toprule
    \textbf{Variant} & \textbf{ASR} & \textbf{CCS} & \textbf{CBF\%} & \textbf{BSR} & \textbf{Rnd} \\
    \midrule
    ZERO-APT (full) & 79\% & 0.860 & 10.3\% & 0.661 & 3.96 \\
    $-$ActionLib & 48\% & 0.510 & 29.9\% & 0.310 & 5.18 \\
    $-$MultiReAct & 42\% & 0.420 & 28.5\% & 0.188 & 6.22 \\
    \bottomrule
  \end{tabular*}
\end{table}

Stripping the action library cuts CCS from 0.860 to 0.510 ($-$40.7\%, $p<0.001$). Without annotated attack knowledge and mechanical precondition checks, LLM-generated commands lose causal structure and CBF\% rises from 10.3\% to 29.9\%. The consistency loss then compounds into an exposure penalty, as the extra blind exploration under continuous Defender monitoring leaks attack intent and steadily prunes the remaining paths.

\begin{figure}[t]
  \centering
  \includegraphics[width=\columnwidth]{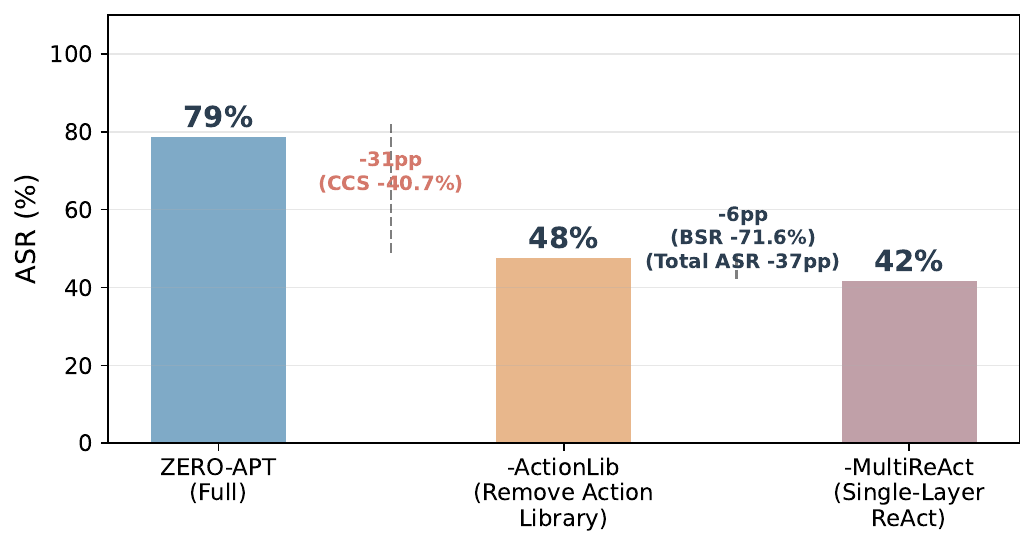}
  \caption{Architecture ablation waterfall chart. From ZERO-APT's 79\%, -ActionLib drops to 48\%, -MultiReAct drops to 42\%; independent component contributions annotated.}
  \label{fig:ablation_waterfall}
\end{figure}

Collapsing the multi-layer architecture cuts ASR from 79\% to 42\% ($-$37pp) and BSR from 0.661 to 0.188 ($-$71.6\%, $p<0.001$). Under adversarial pressure, knowing when to switch strategies (Planner backtracking with Judge feedback) matters more than knowing what strategies are available (action library). Once the three layers fold into a single ReAct loop and Judge closure is gone, the system can still adjust commands within a round but no longer revisits the overall path from cross-round detection results. The ablation conflates multi-layer hierarchy with Judge closure; a full factorial study is left to future work. Figure~\ref{fig:ablation_waterfall} visualizes the per-component drop.

\subsection{Auditability: A Case Study}

One CTI report fragment from a successful S2 run (ASR=1, CCS=0.91, four rounds) makes the audit trail concrete and shows the two backtracking mechanisms in action under real adversarial pressure.

\noindent\colorbox{gray!15}{\parbox{\dimexpr\columnwidth-2\fboxsep\relax}{%
\textbf{R1: Hard-constraint downgrade.} An attempted \texttt{HKLM\textbackslash...\textbackslash Run} write fails with \texttt{ERROR\_ACCESS\_DENIED}; the Executor falls back to \texttt{HKCU} within the same round under its parameter-retry loop. The Defender flags the registry modification (T1547.001, EID 13) and the Judge rules Blue Win, but the reconnaissance findings survive the ruling, a deliberate design choice that preserves information continuity across rounds.

\textbf{R2: Passive backtracking (Judge-driven).} The Planner reads R1's observation and switches to a scheduled task (T1053.005). Both techniques share the precondition of a writable user-level persistence location in the action library Directed Acyclic Graph (DAG), so switching costs nothing. (T1547.001 was preferred initially on minimum-action-cost grounds, with one command versus three for T1053.005.) \texttt{schtasks /create} succeeds undetected.

\textbf{R3$\rightarrow$R4: Tactical switch.} A UAC bypass via cmstp.exe (T1548.002) is blocked by anomalous parent-child detection. The Planner judges all Living Off the Land Binary (LOLBin) variants to have $>$80\% detection probability and switches directly to token manipulation (T1134.002), which the action library metadata marks as the action with the fewest Sysmon events. After verifying that a SYSTEM process exists via \texttt{tasklist /SVC}, the Attacker escalates to \texttt{nt authority\textbackslash system} via \texttt{ImpersonateNamedPipeClient} on a controlled named pipe.
}}

\noindent The case exposes three audit layers. Every strategy adjustment carries a clear trigger and outcome. Defender detections and Planner reasoning sit side by side, opening up bidirectional attack-defense review. The two backtracks, hard-constraint-driven within a round and detection-driven across rounds, are annotated separately. A monolithic ReAct system leaves no comparable trace, so causal breaks resist post-run diagnosis. To gauge whether the Judge's audit reaches a usable bar, we sample 50 CTI reports and ask two human security experts to assess them independently along the same three layers; the resulting ICC(A,1) is 0.81 with 88\% layer-level agreement, on par with the CCS expert scoring reliability reported in Appendix~\ref{sec:metrics} and indicating that the Judge's audit reaches a usable reference standard.

\section{Discussion}
\label{sec:discussion}

\textbf{Defender scope and upgrade paths.} The Defender deliberately stops at Sysmon-based detection~\cite{le2021log,he2024llmelog} without active response. We use intelligent defense in a forward-looking sense, since the current Defender captures adaptive detection rather than fully realized intelligence. Adding techniques such as Cloak, Honey, and Trap~\cite{ayzenshteyn2025cloak} would shift what ASR measures, from strategy adaptability under detection to attack capability against deception. Active deception and multi-agent defense coordination~\cite{xu2025l2m} layer naturally onto the current Defender for richer future regimes.

\textbf{Residual LLM dependency.} The system enforces causal consistency, yet the Planner's reasoning and the Judge's evidence scoring still ride on the LLM. The architecture constrains rather than replaces the LLM, narrowing the surface where errors can emerge without eliminating them.

\textbf{Where hard-constraint filtering turns conservative.} Under near-zero defense pressure (L1), the filter can over-prune. Claude Code edges past ZERO-APT on S2 (20/20 vs.\ 18/20), and both misses trace to a LOLBin path the filter rules out. We still keep filtering on at every level because an attacker cannot read the target's intensity in advance. The L2/L3 net benefit (14pp/15pp) far outweighs the L1 loss (0pp), and L1 itself grows scarcer as intelligent defense moves from option to default.

\textbf{Cross-environment portability.} Section~\ref{sec:realism} settles external validity and internal fairness on our benchmark. To probe cross-environment portability, we strip ZERO-APT down to its Attacker and run it on five HackTheBox machines (Table~\ref{tab:htb}); since HTB carries no intelligent defense, the test reads only as an action-portability check.

\begin{table}[!t]
  \centering
  \caption{ZERO-APT pure Attacker cross-environment validation on HTB.}
  \label{tab:htb}
  \begin{tabular*}{\columnwidth}{@{\extracolsep{\fill}}llcc@{}}
    \toprule
    \textbf{HTB Machine} & \textbf{Difficulty} & \textbf{Attempted} & \textbf{Completed} \\
    \midrule
    Active & Easy & 5 & 5 \\
    Forest & Easy & 4 & 4 \\
    Sauna & Easy & 4 & 3 \\
    Cicada & Easy & 3 & 2 \\
    Certified & Medium & 5 & 2 \\
    \bottomrule
  \end{tabular*}
\end{table}

On HTB the pure Attacker tracks other multi-agent penetration tools~\cite{shen2025pentestagent,xu2024autoattacker}, exactly as expected if ZERO-APT's gains came from architecture rather than the LLM. Closed-loop gaming under intelligent defense remains the primary evaluation target.

\textbf{Future outlook.} Two limitations call for community-built infrastructure: the action library covers only Windows Server 2022 post-exploitation, and the testbed runs on two VMs. The six-tuple action abstraction is cross-platform, and recent work runs LLM agents through autonomous penetration on enterprise-grade Active Directory networks~\cite{happe2025can}. The natural next step is adversarial co-evolution, where the Defender learns from past games while the Attacker adapts in step within an outer loop, supported by hierarchical multi-agent RL~\cite{singh2024hierarchical,tang2024method} and LLM-enhanced multi-agent defense~\cite{xu2025l2m} as templates, with the monotonic ASR decline already serving as a first feasibility signal.

\section{Ethical Considerations}
\label{sec:ethics}

Every attack action originates from publicly available open-source tools (Atomic Red Team) and MITRE ATT\&CK techniques, and every experiment runs inside an isolated VM with no inbound or outbound traffic to live networks. We will release the codebase, experiment code, and scenario configurations under a responsible-use declaration that limits use to authorized security testing, academic research, and defense validation. The configurable Defender component is designed to lower the barrier for defenders to evaluate and harden their own SOC pipelines against LLM-driven adversaries.

\section{Conclusion}
\label{sec:conclusion}

This work tackles three concerns that LLM-driven penetration agents face under intelligent defense: realism, causal consistency, and auditability. ZERO-APT addresses them inside a single attacker-defender-judge framework, with planning-execution separation, multi-dimensional ReAct feedback, and a hard-constraint-filtered action library lifting causal consistency out of unreliable LLM reasoning, and an independent Judge that adjudicates each round and emits structured CTI reports for end-to-end auditability. Across five scenarios and three Defender levels, the architectural advantage widens with defense intensity, and ablation traces the gains to architecture rather than raw LLM capability. Scaling LLMs alone does not close these gaps under live defense; architecture-level mechanisms do. We release the codebase and scenario configurations as a benchmark for attacker-defender co-evolution.

\bibliographystyle{IEEEtran}
\bibliography{refs}

\appendices
\section{Evaluation Metrics Definition}
\label{sec:metrics}

We use 13 evaluation metrics organized across three categories. The four core metrics---ASR, CCS, CBF\%, and BSR---are defined in Section~\ref{sec:evaluation}. Table~\ref{tab:metrics_full} provides complete definitions for all 13 metrics, including auxiliary measures referenced throughout the paper.

\begin{table*}[t]
  \centering
  \caption{Complete metric definitions.}
  \label{tab:metrics_full}
  \renewcommand{\arraystretch}{1.25}
  \begin{tabular*}{\textwidth}{@{\extracolsep{\fill}}c>{\raggedright\arraybackslash}p{1.4cm}>{\raggedright\arraybackslash}p{3.5cm}>{\raggedright\arraybackslash}p{5.5cm}>{\raggedright\arraybackslash}p{2.2cm}@{}}
    \toprule
    \textbf{\#} & \textbf{Abbr.} & \textbf{Full Name} & \textbf{Definition} & \textbf{Formula} \\
    \midrule
    1 & ASR & Attack Success Rate & Binary: 1 if Attacker achieves mission goal within the round limit, 0 otherwise & ASR = N\_success / N\_total \\
    2 & Time & Total Execution Time & Wall-clock seconds from experiment start to termination (PentestGPT excludes human interaction delay) & --- \\
    3 & CCS & Causal Consistency Score & Two independent human experts review each complete attack chain, deducting 1 point per causal inconsistency from a starting score of 10, averaging and normalizing to $[0,1]$. ICC=0.84. Unlike mechanical precondition matching, expert scoring captures implicit causal breaks where preconditions are formally satisfied but logically unsupported & CCS = mean(expert\_scores) / 10 \\
    4 & CBF\% & Causal Break Failure Rate & Fraction of commands that fail (non-zero exit code) due to unmet causal preconditions, excluding commands that succeed but are caught by Defender & CBF\% = N\_failed / N\_total\_commands \\
    5 & BSR & Backtrack Success Rate & Fraction of Planner backtrack events (triggered by detection or precondition failure) that result in successful path recovery & BSR = N\_successful\_backtracks / N\_total\_backtracks \\
    6 & Rpt\% & Repeat Attempt Rate & Fraction of execution steps that repeat the same \texttt{action\_id} within one experiment & Rpt\% = N\_repeat\_steps / N\_total\_steps \\
    7 & TP\% & True Positive Rate & Fraction of rounds where the Defender correctly detects an intrusion & TP\% = N\_detected\_rounds / N\_total\_rounds \\
    8 & FP\% & False Positive Rate & Fraction of rounds where the Defender issues a false alert & FP\% = N\_fp\_rounds / N\_total\_rounds \\
    9 & Rnd & Attack Chain Length & Number of game rounds from experiment start to goal achievement or termination & --- \\
    10 & ActD & Action Diversity & Number of distinct \texttt{action\_id}s used in one experiment & --- \\
    11 & HCT & Hard Constraint Triggers & Number of candidate actions excluded by the Dispatcher due to mechanical precondition mismatch & --- \\
    12 & CoT & CoT Coverage & Fraction of Planner and Judge decision steps with complete chain-of-thought records & CoT = N\_steps\_with\_CoT / N\_total\_decision\_steps \\
    13 & ICC & Inter-Rater Reliability & Intraclass correlation coefficient for CCS expert scoring: ICC(A,1)=0.84, 95\% CI [0.78, 0.89] & ICC(A,1) two-way mixed \\
    \bottomrule
  \end{tabular*}
\end{table*}

\textbf{Cross-scheme metric comparability.} The following metrics do not compare across schemes due to architectural differences: Time is inapplicable to Aurora (static PDDL script execution) and PentestGPT (includes human interaction delay); HCT and CoT apply only to ZERO-APT (other schemes lack runtime action library filtering or CoT tracking); BSR carries different semantics across systems (ZERO-APT measures Planner subgoal-level backtracking, while baselines measure command-level retries). These incomparabilities reflect genuine architectural differences rather than experimental design flaws.

\section{Scenario Specifications}
\label{sec:scenarios_appendix}

Table~\ref{tab:scenarios_full} gives complete configurations for the five adversarial scenarios.

\begin{table*}[t]
  \centering
  \caption{Scenario specifications.}
  \label{tab:scenarios_full}
  \begin{tabular*}{\textwidth}{@{\extracolsep{\fill}}p{1.8cm}p{2.2cm}p{2.2cm}p{2.2cm}p{2.2cm}p{2.2cm}@{}}
    \toprule
    & \textbf{S1: Credential Theft} & \textbf{S2: Persistence + Privesc} & \textbf{S3: Lateral Movement} & \textbf{S4: Covert Exfiltration} & \textbf{S5: Path Blocking} \\
    \midrule
    \textbf{Primary Target} & ZEROAPTtarget (192.168.32.139) & ZEROAPTtarget (192.168.32.139) & ZEROAPTtarget + DataServer (192.168.1.102) & ZEROAPTtarget (192.168.32.139) & ZEROAPTtarget (192.168.32.139) \\
    \textbf{Defender Level} & L2 & L2 & L3 & L2 & L1 \\
    \textbf{Initial Access} & User-level shell (User1) & User-level shell (User1) & User-level shell (User1) on both hosts & User-level shell (User1) & User-level shell (User1), primary path pre-blocked \\
    \textbf{Mission Goal} & Extract local credentials & Establish persistence + escalate to SYSTEM & Move laterally to DataServer + collect target file & Discover, collect, exfiltrate sensitive data & Achieve SYSTEM privileges via alternative path \\
    \textbf{Tested Capability} & Basic multi-step causal chain execution & Multi-branch path selection under hard constraints & Adaptive bypass under maximum detection pressure & End-to-end long-chain execution stability & Backtrack recovery when primary path blocked \\
    \textbf{Key Adversarial Feature} & Low Defender detection (routine discovery) & Medium detection (monitored techniques) & Maximum detection (L3 full TTP); cross-subnet movement & Stealth-required: detection = mission compromise & Primary path pre-blocked; Attacker must discover and reroute \\
    \bottomrule
  \end{tabular*}
\end{table*}

\textbf{Scenario design logic.} The five scenarios stress three capabilities in turn. Causal consistency is tested at increasing difficulty: S1 targets a basic chain, S2 adds multi-branch path selection, and S4 extends to a four-stage stealth chain. Adaptive bypass is isolated in S3 under L3 detection with cross-subnet lateral movement. Backtrack recovery is evaluated in S5, where the primary path is pre-blocked to isolate environmental constraints from detection pressure. Section~\ref{sec:realism} Table~\ref{tab:per_scenario_asr} reports the per-scenario ASR breakdown.

\section{Action Library Structure}
\label{sec:action_lib}

The action library contains 775 post-exploitation actions derived from Atomic Red Team, covering 12 MITRE ATT\&CK tactics and 193 distinct techniques. Each action follows the six-tuple structure defined in the main text Table~\ref{tab:action_structure}. Hard conditions (privilege levels, tool dependencies, port availability) are defined in structured fields, mechanically matched by the Dispatcher against environmental state; soft conditions (specific file paths, account names, configuration values) are dynamically acquired by the Executor through exploration commands and written to the environment knowledge base. Current action library coverage: Discovery (TA0007), Persistence (TA0003), Privilege Escalation (TA0004), Defense Evasion (TA0005), Credential Access (TA0006), Lateral Movement (TA0008), Collection (TA0009), Command and Control (TA0011), Exfiltration (TA0010), Execution (TA0002), Impact (TA0040). All actions target Windows.

\section{Defender Detection Rules}
\label{sec:defender_rules}

\textbf{L1 baseline (noise + no intelligence).} Sysmon logs mixed with normal Windows activity noise (file browsing, service start/stop, Windows Update, routine PowerShell). No signature rules, no threat intelligence. Detection relies purely on Monitor LLM anomaly detection against background noise.

\textbf{L2 standard (clean logs + no intelligence).} Clean Sysmon logs containing Attacker activity only. Four Sysmon signature rules (see Section~\ref{sec:defender}): EID 1 (suspicious parent-child processes), EID 12/13/14 (persistence-related registry modifications), EID 11 (executable creation in system/startup directories), EID 3 (non-standard/C2 port connections). Monitor LLM performs contextual reasoning and false-positive suppression on rule-matched events. Full prompt template in open-source repository.

\textbf{L3 enhanced (clean logs + intelligence).} L2 rules plus Attacker identity injection (``LLM Agent known to use MITRE ATT\&CK post-exploitation techniques''), full action library technique list, and accumulated detection history.

\section{CTI Report Example}
\label{sec:cti_example}

The following is an abridged CTI report generated by ZERO-APT's Judge for one successful S2 (Persistence + Privilege Escalation) experiment (exp\_20260511\_190000\_S2, ASR=1, CCS=0.91, 4 rounds). The full CTI report (approximately 2,500 words, covering the complete experimental timeline, the evolution of forbidden actions, hard-constraint triggers, red-team decision highlights, and defensive improvement recommendations) is available in the open-source repository.

\begin{table*}[!ht]
\centering
\colorbox{gray!15}{\parbox{\dimexpr\textwidth-2\fboxsep-6pt\relax}{%
\textbf{Experiment Summary.}\vspace{2pt}

\begin{itemize}[leftmargin=*,nosep]
\item Scenario: S2 (Persistence + Privilege Escalation)
\item Mission: Establish persistence on ZEROAPTtarget and escalate to SYSTEM
\item Result: 4 rounds, success
\item Blue Win rounds: 2/4 (R1, R3); TP Rate: 2/4 (50.0\%)
\end{itemize}

\vspace{4pt}
\textbf{Final Attack Chain (abridged).}\vspace{2pt}

\footnotesize
\renewcommand{\arraystretch}{1.25}
\begin{tabular*}{\dimexpr\textwidth-2\fboxsep-12pt\relax}{@{\extracolsep{\fill}}c>{\raggedright\arraybackslash}p{4cm}>{\raggedright\arraybackslash}p{2cm}>{\raggedright\arraybackslash}p{6.5cm}@{}}
\toprule
\textbf{R} & \textbf{Action} & \textbf{Technique} & \textbf{Result} \\
\midrule
R1 & Recon: \texttt{whoami /groups}, \texttt{systeminfo}, file permission check $\rightarrow$ attempted \texttt{HKLM{\char`\\}...{\char`\\}Run} write failed (ERROR\_ACCESS\_DENIED) $\rightarrow$ downgraded to HKCU, executed \texttt{reg add HKCU{\char`\\}...{\char`\\}Run} & T1087, T1082, T1083; T1547.001 & Reconnaissance undetected; T1547.001 blocked by Defender (Sysmon EID 13, confidence 85\%). Note: Executor wrote recon discoveries to global state before Judge ruling; Blue Win does not revert already-written discoveries. \\
R2 & Scheduled task persistence via \texttt{schtasks /create} (R1$\rightarrow$R2 cross-round backtrack: Planner reads R1 observation then switches) & T1053.005 & Persistence successful (undetected) \\
R3 & UAC bypass via \texttt{cmstp.exe /s bypass.sct} & T1548.002 & Blocked by Defender (anomalous parent-child, confidence 78\%) \\
R4 & \texttt{tasklist /SVC} confirms SYSTEM process exists $\rightarrow$ token manipulation via named pipe impersonation $\rightarrow$ \texttt{whoami} verification & T1057, T1134.002, T1059.003 & Escalation to SYSTEM successful (undetected) \\
\bottomrule
\end{tabular*}

\vspace{4pt}
\textbf{Adversarial Dynamics.}\vspace{2pt}

\textit{Hard-constraint-driven downgrade (R1):} \texttt{HKLM{\char`\\}...{\char`\\}Run} write fails with \texttt{ERROR\_ACCESS\_DENIED}; the Executor falls back to HKCU within the same round under its parameter-retry loop, a hard-constraint-driven adjustment independent of Defender detection.

\textit{Passive backtracking (R1$\rightarrow$R2, Judge-driven):} Defender detects T1547.001; Judge rules Blue Win; R2 Planner reads observation and switches to T1053.005 with zero additional reconnaissance cost, since both techniques share a writable user-level persistence location in the action library DAG.

\textit{Passive backtracking (R3$\rightarrow$R4, Judge-driven):} T1548.002 blocked by anomalous parent-child detection. Planner determines alternative LOLBin variants have $>$80\% detection probability, switches to token manipulation (T1134.002). Verifies SYSTEM process existence via \texttt{tasklist /SVC}, escalates through named pipe impersonation.

\vspace{4pt}
\textbf{Planner Decision Highlights.}\vspace{2pt}

\begin{itemize}[leftmargin=*,nosep]
\item R1: combined reconnaissance and first-attempt strategy; Executor writes discoveries before Judge ruling, preserving knowledge for subsequent switching.
\item R2: T1547.001 preferred over T1053.005 on minimum-action-cost grounds (one command vs.\ three).
\item R3$\rightarrow$R4: tactical switch references action library metadata: T1134.002 produces the fewest Sysmon events; PipeEvent monitoring (EID 17/18) is typically disabled under L2.
\end{itemize}
}}
\end{table*}

\section{Statistical Test Details}
\label{sec:statistics}

Table~\ref{tab:stats} provides complete statistical results for all hypothesis tests in Section~\ref{sec:evaluation}. All tests use two-sided alternatives. Chi-squared tests (2$\times$2 tables) use Yates continuity correction~\cite{yates1934contingency}. Welch's $t$-tests~\cite{welch1947generalization} assume unequal variances. Wilcoxon signed-rank tests~\cite{wilcoxon1945individual} are used for Holm-corrected post-hoc pairwise comparisons. Page's $L$ test~\cite{page1963ordered} evaluates linear ASR trend across Defender intensities. Friedman test~\cite{friedman1937use} evaluates overall differences among repeated measures.

\begin{table*}[t]
  \centering
  \caption{Complete statistical test results.}
  \label{tab:stats}
  \begin{tabular*}{\textwidth}{@{\extracolsep{\fill}}lp{2.6cm}cccccc@{}}
    \toprule
    \textbf{Hypothesis} & \textbf{Test} & \textbf{Statistic} & \textbf{df} & \(p\) & \textbf{Effect Size} & \textbf{95\% CI} & \textbf{Sig.\ (\(\alpha{=}0.05\))} \\
    \midrule
    H1: ASR\_ZA $>$ ASR\_Aurora & $\chi^2$ (Yates) & 16.43 & 1 & $<$0.001 & $\phi$=0.74 & --- & Yes \\
    H1: ASR\_ZA $>$ ASR\_PGPT & $\chi^2$ (Yates) & 7.04 & 1 & 0.008 & $\phi$=0.48 & --- & Yes \\
    H1: ASR\_ZA $>$ ASR\_CC & $\chi^2$ (Yates) & 1.68 & 1 & 0.195 & $\phi$=0.24 & --- & No \\
    H2: CCS\_Aurora $>$ CCS\_ZA & Welch's $t$ & 2.63 & 27.8 & 0.014 & $d$=0.96 & [0.009, 0.075] & Yes \\
    H2: CCS\_ZA $>$ CCS\_PGPT & Welch's $t$ & 27.38 & 27.4 & $<$0.001 & $d$=10.1 & [0.445, 0.517] & Yes \\
    H2: CCS\_ZA $>$ CCS\_CC & Welch's $t$ & 19.35 & 27.2 & $<$0.001 & $d$=7.06 & [0.310, 0.384] & Yes \\
    H3: Time\_ZA $<$ Time\_CC (overall) & Welch's $t$ & $-$0.93 & 26.8 & 0.361 & $d$=$-$0.36 & [$-$411.4, 158.4] & No \\
    H3: Time\_ZA $<$ Time\_CC (L2/L3) & Welch's $t$ & $-$4.12 & 21.6 & 0.003 & $d$=$-$1.69 & --- & Yes \\
    H4: CBF\%\_ZA $<$ CBF\%\_PGPT & Welch's $t$ & $-$17.83 & 20.2 & $<$0.001 & $d$=$-$6.54 & [$-$0.311, $-$0.247] & Yes \\
    H4: CBF\%\_ZA $<$ CBF\%\_CC & Welch's $t$ & $-$10.60 & 27.6 & $<$0.001 & $d$=$-$4.04 & [$-$0.203, $-$0.137] & Yes \\
    H5: ASR monotonic (scenarios) & Friedman & $\chi^2$=8.40 & 2 & 0.015 & --- & --- & Yes \\
    H5: ASR linear trend & Page's $L$ & $L$=118 & --- & $<$0.01 & --- & --- & Yes \\
    H6: CCS\_ZA $>$ CCS\_{-ActionLib} & Welch's $t$ & 9.12 & 27.3 & $<$0.001 & $d$=3.36 & [0.277, 0.439] & Yes \\
    H7: BSR\_ZA $>$ BSR\_{-MultiReAct} & Welch's $t$ & 5.87 & 18.5 & $<$0.001 & $d$=2.52 & [0.324, 0.720] & Yes \\
    \bottomrule
  \end{tabular*}
\end{table*}

\textbf{CCS inter-rater reliability.} ICC(A,1)=0.84, CI [0.78, 0.89]. CCS expert scoring captures implicit causal breaks beyond mechanical precondition matching.

\textbf{Effect size interpretation.} Cohen's $d$: small (0.2), medium (0.5), large (0.8). All significant comparisons show medium to very large effects ($d > 2.0$ for H2 vs.\ PentestGPT and Claude Code; $d > 3.0$ for H6). Holm-Bonferroni correction across the 12 pairwise comparisons leaves all originally significant results significant at $\alpha=0.05$ (the H1 ASR\_ZA vs.\ CC comparison remains non-significant).

\section{LLM Module Input/Output Formats}
\label{sec:llm_formats}

ZERO-APT uses 7 LLM modules connected through the per-round game loop. All input/output formats are defined in prompt templates (\texttt{agents/prompts/*.py}) and enforced at runtime by JSON schema validation. Per round: Planner $\rightarrow$ Dispatcher-P1 $\rightarrow$ Dispatcher-P2 $\rightarrow$ Executor. The Executor output is submitted to the Judge alongside the Defender Monitor output; the Judge adjudicates the round and updates global state, and its feedback returns to the Planner in the next round. After the experiment terminates, the round logs feed the Summary module, which produces a structured CTI report (full example in Section~\ref{sec:cti_example}). Table~\ref{tab:llm_schema} lists the input/output fields of all seven modules.

\begin{table*}[t]
  \centering
  \caption{LLM module input/output schemas (enforced at runtime by JSON schema validation).}
  \label{tab:llm_schema}
  \renewcommand{\arraystretch}{1.15}
  \begin{tabular*}{\textwidth}{@{\extracolsep{\fill}}lp{5.4cm}p{5.4cm}p{2.4cm}@{}}
    \toprule
    \textbf{Module} & \textbf{Input fields} & \textbf{Output fields} & \textbf{Schema} \\
    \midrule
    Planner & \texttt{mission, environment, discovered\_info, observation, plan, current\_subgoal, forbidden\_actions, pending\_feedback} & \texttt{plan, current\_subgoal, reasoning, goal\_achieved} & JSON, strict \\
    Dispatcher-P1 & \texttt{plan, subgoal, action\_lib\_str, archive\_environment, discovered\_info, forbidden\_actions} & \texttt{selected\_actions [\{action\_id, reasoning\}], needs\_feedback, feedback\_reason} & JSON, strict \\
    Dispatcher-P2 & \texttt{subgoal, selected\_actions\_str, discovered\_info, archive\_environment} & \texttt{AttackPlan [\{stepN, reasoning, max\_retries\}]} & JSON, strict \\
    Executor & \texttt{mini\_plan, step\_id, executed\_commands, local\_environment\_cache, current\_credentials, discovered\_info} & \texttt{thought, action, command, executor, discoveries, analysis\_feedback} & JSON, strict \\
    Defender Monitor & Sysmon events from ELK \texttt{winlogbeat-*} (1-minute sliding window), optional TTP intelligence at L3 & \texttt{has\_intrusion, detected\_techniques, confidence (0--100), evidence, analysis} & JSON, strict \\
    Judge & \texttt{red\_feedback, blue\_feedback, archive\_environment, discovered\_info, forbidden\_actions, target\_info} & \texttt{blue\_evidence\_score (0--100), winner, observation, environment\_update} & JSON, strict \\
    Summary & \texttt{mission, round\_logs} & 7-section structured CTI report (overview, timeline, strategy evolution, detection effectiveness, forbidden-action evolution, defensive recommendations, traceable decision chain) & Markdown, sectioned \\
    \bottomrule
  \end{tabular*}
\end{table*}

\section{LLM Usage Statement}
\label{sec:llm_usage}

LLMs sit at the core of ZERO-APT's methodology, with all seven agent modules relying on LLM reasoning as detailed in Sections~\ref{sec:attacker}--\ref{sec:judge}. Below we supplement configuration and resource details, and disclose LLM use in manuscript preparation.

\textbf{LLM backend.} All seven agent modules use Claude Opus 4.7, which is Anthropic's latest Claude release at the time of writing, with each module bound to a separate API key for cross-agent information isolation, default temperature (1.0), no max\_tokens limit, and outputs enforced by JSON schema validation. Hosted-API access pins the model version for reproducibility over a substantial deployment window. Ablation variants and baselines use matching backends.

\textbf{Token consumption and cost.} $\sim$55--66K tokens/round ($\sim$\$0.40/round), varying by task and detection depth.

\textbf{Editorial use.} LLMs were used for editorial purposes, specifically for aligning LaTeX formatting and fine-tuning figure-generation scripts. All outputs were inspected by the authors to ensure accuracy and originality.

\textbf{Per-scenario round limits.} Maximum round limits are set per scenario (S1=8, S2=10, S3=12, S4=12, S5=9) based on empirical observation during pilot experiments. These limits are set well above the maximum observed rounds in each scenario to avoid artificially truncating successful runs while preventing infinite loops. The generous caps also ensure that all three baselines have sufficient rounds to produce attack chains long enough for meaningful causal consistency analysis.

\textbf{Defender detection window.} The Defender Monitor queries Sysmon events from the ELK Stack within a 1-minute sliding window each round. The window size is configurable, but all experiments in this paper use a fixed 1-minute window to ensure fair comparison across schemes and scenarios. Shorter windows may miss late-arriving events, while longer windows introduce latency without material detection gain at the current configuration.

\end{document}